\documentclass[prb,twocolumn,superscriptaddress]{revtex4-1}

\usepackage[USenglish]{babel} 
\usepackage{amsmath}  
\usepackage{amsfonts} 
\usepackage{graphicx} 
\usepackage{color}
\usepackage[export]{adjustbox} 
\usepackage{comment} 
\usepackage{natbib} 

\begin{document}
\title{Illustrations of loosely-bound and resonant states in atomic nuclei}

\author{A. C. Dassie}
\affiliation{Department of Physics FCEIA (UNR),
             Av. Pellegrini 250, S2000BTP Rosario, Argentina}
\affiliation{Institute of Nuclear Studies and Ionizing Radiations (UNR), 
		    	Riobamba y Berutti, S2000EKA Rosario, Argentina.}             
\affiliation{Physics Institute of Rosario (CONICET-UNR), 
             Esmeralda y Ocampo, S2000EZP Rosario, Argentina}

\author{F. Gerdau}
\affiliation{Department of Physics FCEIA (UNR),
             Av. Pellegrini 250, S2000BTP Rosario, Argentina}
\affiliation{Institute of Nuclear Studies and Ionizing Radiations (UNR), 
		    	Riobamba y Berutti, S2000EKA Rosario, Argentina.}             

\author{F. J. Gonzalez}
\affiliation{Department of Physics FCEIA (UNR),
             Av. Pellegrini 250, S2000BTP Rosario, Argentina}
\affiliation{Institute of Nuclear Studies and Ionizing Radiations (UNR), 
		    	Riobamba y Berutti, S2000EKA Rosario, Argentina.}             

\author{M. Moyano}
\affiliation{Department of Physics FCEIA (UNR),
             Av. Pellegrini 250, S2000BTP Rosario, Argentina}
\affiliation{Institute of Nuclear Studies and Ionizing Radiations (UNR), 
		    	Riobamba y Berutti, S2000EKA Rosario, Argentina.}             

\author{R. M. Id Betan}
\affiliation{Department of Physics FCEIA (UNR),
             Av. Pellegrini 250, S2000BTP Rosario, Argentina}
\affiliation{Institute of Nuclear Studies and Ionizing Radiations (UNR), 
		    	Riobamba y Berutti, S2000EKA Rosario, Argentina.}             
\affiliation{Physics Institute of Rosario (CONICET-UNR), 
             Esmeralda y Ocampo, S2000EZP Rosario, Argentina}

\date{\today}

%
%

\begin{abstract}
Using the one-dimensional potential well with realistic parameters for atomic nuclei, we illustrate the movement of the poles of the $S$-matrix, and the transmission coefficient when the well supports an anti-bound state. We calculate the phase shift of the atomic nuclei $^5$He using the three-dimensional potential well, and compare it with the experimental one. The paper gives a glance of some of the properties found in realistic loosely and resonant nuclear systems, using equations at the level of undergraduate students.
\end{abstract}

\maketitle

\section{Introduction} 

Resonance is important in both classical and quantum mechanics. In classical mechanics, resonance is associated with an increase in amplitude when the frequency of an external force coincides with a natural frequency of the system. In quantum mechanics, resonances are often defined as peaks of the transmission coefficient. They are also defined as 
isolated poles of the $S$-matrix, or as rapid variations of the phase shift. In this article we explore these definitions of resonances using the analytic solutions of the one and three-dimensional potential well, and we show how this model can be used to illustrate some of the properties of atomic nuclei. 

We begin by reviewing resonances in the familiar damped harmonic oscillator (Section \ref{sec.ho}), and then consider the well known bound and scattering solutions of the one dimensional potential well in Sections \ref{sec.bound} and \ref{sec.scatt}. The less known solution of the problem using the $S$-matrix formalism is developed in Section \ref{sec.smatrix}. In order to present resonances in the three-dimensional space we illustrate the phase shift in Section \ref{sec.phase} using the solution of the three-dimensional potential well.

In Section \ref{sec.moving} we illustrate how the different kinds of poles of the $S$-matrix evolve from resonances to bound states for the one-dimensional potential well. In this section we also compare the poles of the $S$-matrix with the peaks of the transmission coefficient. In Section \ref{sec.deuteron} we use the one-dimensional potential well as a toy model to show the characteristic behavior of the loosely bound state wave function of the deuteron. We also connect some of the poles of the previous section with the bound and anti-bound states of the deuteron. In the second part of Section \ref{sec.deuteron} we illustrate the characteristic behavior of the resonant phase shift in the $^5$He nucleus using the three-dimensional potential well. Finally, in Sec. \ref{sec.conclusions} we summarize our results.

\section{Damped harmonic oscillator}\label{sec.ho}
We begin by reviewing the well-known solution to the classical damped harmonic oscillator so that in the following sections we can point out some commonalities with the quantum mechanical problem. The equation of motion for a particle of mass $m$ subject to a driving force $F(t)$ is 
\begin{equation}\label{eqn:oscilador}
   \frac{d^2 x}{dt^2} + 2\gamma \frac{dx}{dt} + \omega^2_{0}x = \frac{F(t)}{m} = f(t)\, ,
\end{equation}
where $\omega_{0}$ and $\gamma$ are the natural frequency and the damping coefficient of the system, respectively, and both have positive values.

With an oscillating driving force $f(t) = F_{\omega} e^{-i\omega t}$, the solution is
\begin{equation}
    x(t) = X_{\omega}e^{-i\omega t} ,
\end{equation}
with
\begin{equation}\label{eqn:xgf}
    X_{\omega} = -\frac{F_{\omega}}{\omega^2+2i\gamma \omega - \omega_{0}^2} 
    = F_\omega G(\omega) \, .
\end{equation}

We are interested in the properties of the amplitude $G(\omega)$ of the response function because in this paper we will make connections between it and quantum scattering theory. First, we write $G(\omega)$ as follows  
\begin{equation}
    G(\omega) = -\frac{1}{(\omega-\omega_{1})(\omega-\omega_{2})} \, ,
\end{equation}
with
\begin{equation}\label{eqn:omegas}
    \omega_{1,2} = \pm (\omega_{0}^{2} - \gamma^2)^{1/2} - i\gamma .
\end{equation}

\begin{figure}[h!t]
  \centering
  \includegraphics[width=0.8\linewidth,keepaspectratio]{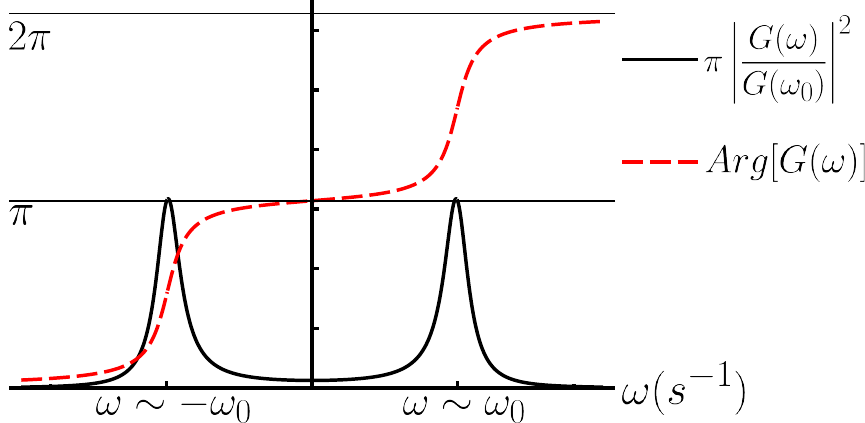}
  \caption{Profile of $\pi |G(\omega)|^2/|G(\omega_0)|^2$ and $Arg[G(\omega)]$ as function of $\omega$ for $\gamma = 0.1\omega_{0}$.}
  \label{fig:polos}
\end{figure}

The complex poles of $G(\omega)$ imprint their signature in the shape of $|G(\omega)|$ and  also in $Arg[G(\omega)]$. For a weak damping, i.e. $\gamma \ll \omega_0$, $|G(\omega)|^2$ has two resonant peaks with width $2\gamma$ centered at $\omega \approx \pm \omega_0$, while the function $Arg[G(\omega)]$ changes by $\pi$, centered at $\pi/2$ and $3\pi/2$, in a narrow range around each one of the resonant frequencies. These characteristics are shown in Fig. \ref{fig:polos} for $\gamma = 0.1\omega_{0}$. Then we can say that a classical resonance is characterized by two isolated poles close to the real axis.

For $\omega \approx \omega_{0}$ we may write
\begin{align}
	\left | \frac{G(\omega)}{G(\omega_0)} \right |^2
		&\approx  \frac{\gamma^2}{(\omega-\omega_{0})^2+\gamma^2} \; ,  \label{eq.bw} \\
   Arg[G(\omega)]
   		& \approx
   				\arctan\left(\frac{\gamma}{\omega-\omega_{0}}\right) \, . \label{eq.bw2} 
\end{align}
The first expressions confirms that the width of the peak in $|G(\omega)|^2$ is $2\gamma$, while the second one shows the rapid variation of $Arg[G(\omega)]$ in the vicinity of $\omega \approx \omega_0$.

Classical and quantum resonances show some common characteristics; in particular, this kind of bump also appears in the experimentally-measured resonant scattering cross-section as a function of the energy in nuclear physics. In this context, a lorentzian parametrization like Eq. (\ref{eq.bw}) is called Breit-Wigner distribution  (see for example pp. 23-24 of Ref. \onlinecite{1989Kukulin}). 
This characteristic behavior of $Arg[G(\omega)]$ shown in Fig. \ref{fig:polos} can also be reproduced in a simple quantum-mechanical model such as the one introduced in Sec. IIID.

\begin{figure*}[h!t]
  \centering
  \includegraphics[width=0.25\linewidth,keepaspectratio]{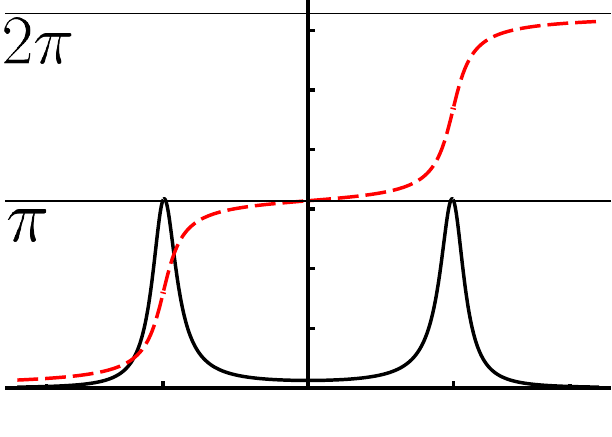}
  \includegraphics[width=0.25\linewidth,keepaspectratio]{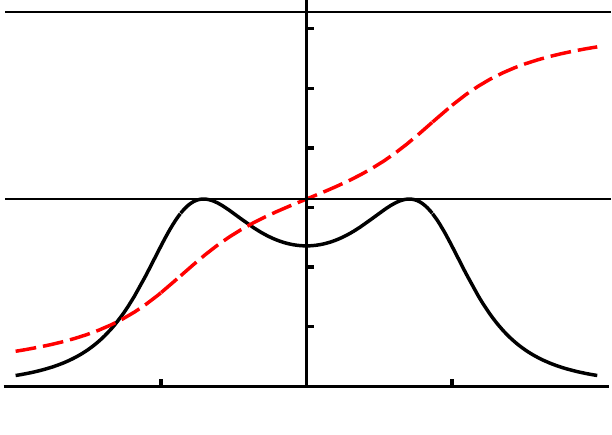}
  \includegraphics[width=0.423\linewidth,keepaspectratio]{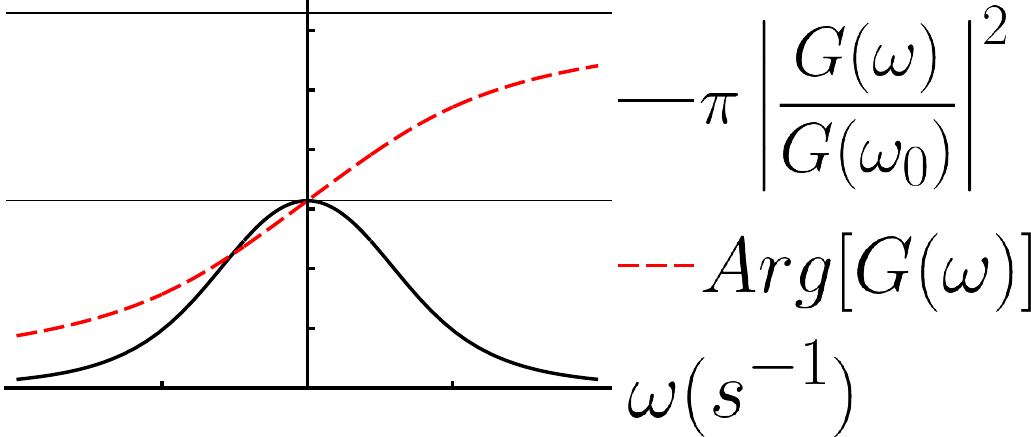}
  \caption{Profile of $\pi |G(\omega)|^2/|G(\omega_0)|^2$ and $Arg[G(\omega)]$ as function of $\omega$ for (left) $\gamma = 0.1\omega_{0}$, (center) $\gamma = 0.5\omega_{0}$, and (right) $\gamma = 0.9\omega_{0}$.}
  \label{fig:evolucion}
\end{figure*}

For the limit $\gamma \rightarrow \omega_{0}$, we find that both poles converge at $\omega_{1}=\omega_{2}=-i \, \omega_0$. So the pole structure of $G(\omega)$ changes from two isolated single poles to a second-order pole at $\omega=-i\omega_{0}$. Figure \ref{fig:evolucion} shows this transition for three values of $\gamma$. We can see the two peaks approach and merge into a single one. We also observe that the argument of $G(\omega)$ changes from two jumps to a single smooth jump of $2\pi$. 

\section{Quantum mechanical scattering}\label{sec.qm}
In this section we consider resonances in quantum mechanics.  In Sections \ref{sec.bound} and \ref{sec.scatt} we obtain the negative and positive-energy solutions of a one-dimensional Schr\"{o}dinger equation for a square potential well. Then in Section \ref{sec.smatrix} we introduce the $S$-matrix for the one-dimensional system. Finally in Section \ref{sec.phase} we consider resonances within the phase shift formalism for a 3-D model.

\subsection{Bound states} \label{sec.bound}
Let us consider two particles of masses $m_1$ and $m_2$ limited to move in one dimension, subject to an attractive interaction $V=-V_0$ ($V_0>0$) when their separation is less than $R$. This problem can be mapped onto a one-dimensional Schr\"{o}dinger equation for a single particle with reduced mass $\mu=m_1 m_2/(m_1+m_2)$:
\begin{equation}
    - \frac{\hbar^2}{2\mu} \frac{d^2 u}{dx^2}
    + V(x) \, u(x) = \varepsilon u(x) \, ,
    \label{eq:schrod}
\end{equation}
with
$$
    V(x)=
    \begin{cases}
        -V_0 & |x| \le R   \\
        0 & |x| > R  \, .
    \end{cases}
$$

We look for the bound state solutions, i.e. $-V_0<\varepsilon<0$, with the usual boundary conditions at $x=\pm R$. With the requirement that $u(x)$ is bounded, we find discrete energies, $\varepsilon_n<0$, that are obtained from
\begin{align}
   \rho_n &= K_n \tan(K_n R) \, , \label{eq.even} \\
 -\rho_n &= K_n \cot(K_n R) \, . \label{eq.odd}
\end{align}
Equations (\ref{eq.even}) and (\ref{eq.odd}) correspond to even and odd solutions, respectively. The parameters $\rho_n$ and $K_n$ are related to energy by
\begin{equation} \label{eq.kbound}
\rho_n=\sqrt{-\frac{2\mu}{\hbar^2}\varepsilon_n} \; , \; 
   \hspace{3mm}
   K_n=\sqrt{\frac{2\mu}{\hbar^2}(V_0+\varepsilon_n)} \; .
\end{equation}

\subsection{Scattering states} \label{sec.scatt} 
In this section we solve the Schr\"{o}dinger Eq. (\ref{eq:schrod}) for the eigenvalues $\varepsilon > 0$. In this case the energies are continuous and the solutions can be expressed as a linear combination of plane waves
\begin{equation}
	u(x)=
	\begin{cases}
	Ae^{i k x}+A'e^{-i k x}      & \, x\leq -R \, \\
	Be^{iK x}+Ce^{-iK x} & \, -R<x<R \, \\
	D'e^{i k x}+De^{-i k x}      & \, x\geq R  \, ,
	\end{cases}
	\label{eq:scattsol}
\end{equation}
with 
\begin{equation}\label{eq.kkappa}
k=\sqrt{\frac{2\mu}{\hbar^2} \varepsilon} \; , \; 
\hspace{3mm}
K=\sqrt{\frac{2\mu}{\hbar^2} (V_0+\varepsilon)} \; .
\end{equation}

At this stage, we have to decide whether the particle is coming from the right or left. Either of these two options must give the same physical result; the calculated observables cannot depend on this choice. In the $S$-matrix approach, both options will be considered simultaneously, but for this section let us assume it comes from the left, so that $D=0$. By matching conditions at $x=\pm R$, we obtain the following expressions for the transmission $t=D'/A$ and reflection $r= A'/A$ amplitudes:
\begin{align}
   t(k) &= \frac{e^{-i2R k}}
   				   {\cos(2RK) - i\frac{k^2+K^2}{2 k K} \sin(2RK)} \, , 
   				   \label{eq.t} \\
   & \nonumber \\
  r(k) &= \frac{i \frac{2\mu}{\hbar^2} \frac{V_0}{2 k K} \sin(2RK) \, e^{-i2R k}}
  			      {\cos(2RK) - i\frac{k^2+K^2}{2 k K} \sin(2RK)} \, .
  			      \label{eq.r}
\end{align}

In the rest of this paper, we will consider the $k$ representation instead of the energy $\varepsilon$ representation; i.e., we will consider $k$ as the independent variable, so we rewrite
\begin{equation}
K(k) = \sqrt{ k^2 + \frac{2\mu}{\hbar^2} V_0} \; .
\end{equation}

The transmission coefficient
$T = |t|^2$ is
\begin{equation}
	T 
	=  \left[
	       1
	         +  \left( \frac{2\mu}{\hbar^2} \frac{V_0}{2 k K} \right)^2
	                 \sin^2(2RK)
 	      \right]^{-1}  \, ,
	\label{eq:transcoef}
\end{equation}
where the maximums $T=1$ occur for $2RK=\pi,\, \, 2\pi,\, \cdots$; that is, when there is an integer number of half-wavelengths of $K$ inside the potential well. From a time-dependent picture \cite{1977Hammer,2021Staelens} a wave packet can be formed from plane waves and the maximums may be interpreted as the constructive interference inside the well.

\subsection{$S$-matrix} \label{sec.smatrix} 
In the $S$-matrix approach the scattering process is divided into three stages. In the first one the particles move one towards the other, but they are so far apart that one can neglect their mutual interaction. In the second stage the particles are close enough to interact, while in the final stage they are far away and then they do not interact any more. The operator that transforms the initial stage into the final stage is called $S$-matrix or scattering matrix. In a time-dependent framework the initial and final stages correspond to the remote past and future, respectively. In our time-independent formulation, the $S$-matrix connects the wave functions of the two asymptotic regions: $x< -R$ and $x > R$. In this section we calculate the $S$-matrix for the one-dimensional potential well. 

Let us start by considering the scattering states $\varepsilon>0$ for $|x|\geq R$ for particles incident from the left and right:
\begin{align}
   u_L(x) &=
	\begin{cases}
	e^{i k x}+r e^{-i k x}      & \, x\leq -R \, \\
	te^{i k x}      & \, x\geq R  \, ,
	\end{cases} \, , \label{eq.ul} \\
   u_R(x) &=
	\begin{cases}
	t' e^{-i k x}      & \, x\leq -R \, \\
	e^{-i k x}+r'e^{i k x}      & \, x\geq R  \, .
	\end{cases}   \label{eq.ur}
\end{align}
The coefficients $t$ and $r$ correspond to the transmission and reflection amplitudes in the asymptotic regions $|x|\geq R$ from Section \ref{sec.scatt}, i.e., Eqs. (\ref{eq.t}) and (\ref{eq.r}). On the other hand $t'$ and $r'$ are the corresponding magnitudes for the reverse process, i.e. a particle coming from the right, with $t'=t$ and $r'=r$. 

\begin{figure}[h!t]
  \centering
  \includegraphics[width=0.5\linewidth,keepaspectratio]{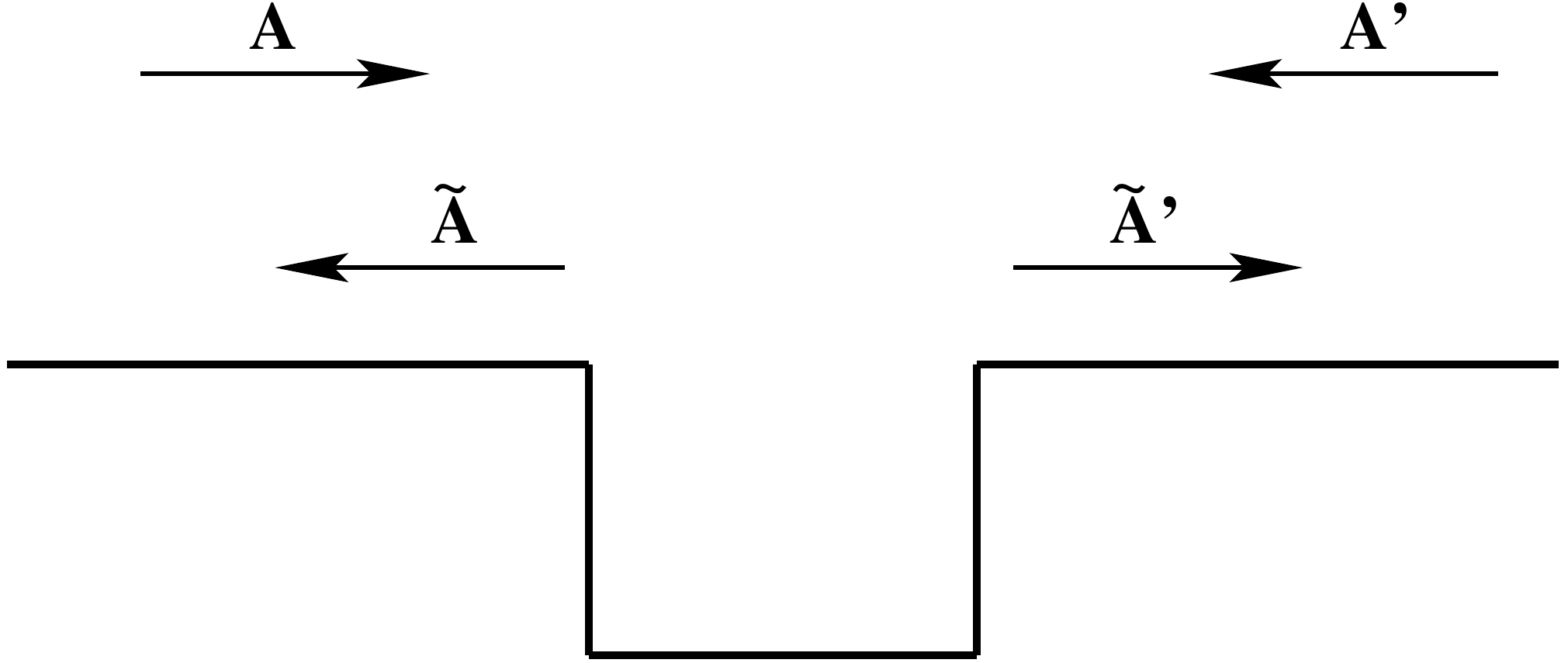}
  \caption{Incoming amplitudes $A$ and $A'$, and outgoing amplitudes $\tilde{A}$ and $\tilde{A}'$ which define the $S$-matrix.}
  \label{fig.Smatrix}
\end{figure}

The general solution is a linear combination of $u_L(x)$ and $u_R(x)$:
\begin{align}
   u(x) &=A u_L(x) + A' u_R(x) \, \\
   &=\begin{cases}
      A e^{i k x}+ \tilde A e^{-i k x}  & \, x\leq -R \\
      \tilde A' e^{i k x} +A'e^{-i k x}   & \, x\geq R  
   \end{cases} \, ,
\end{align}
where $A$ and $A'$ are the amplitudes of the wave function as it approaches to the interaction region, also called incoming amplitudes, while $\tilde{A}$ and $\tilde{A}'$ are the outgoing amplitudes, as shown in Fig. \ref{fig.Smatrix}.

The relationship between the incoming and outgoing amplitudes defines the $S$-matrix:
\begin{equation}\label{eq.s}
  \begin{pmatrix} \tilde A \\ \tilde A' \end{pmatrix}
  		= S
   \begin{pmatrix} A' \\ A \end{pmatrix} \; ,
\end{equation}
and we see from the earlier analysis:
\begin{equation}
	S=\begin{pmatrix} t & r \\ r & t \end{pmatrix} \, .
\end{equation} 

Using Eqs. (\ref{eq.t}) and (\ref{eq.r}) for the transmission $t(k)$ and reflection $r(k)$ amplitudes, respectively, we can write the $S$-matrix as
\begin{widetext}
\begin{equation} \label{eq.smatrix}
   S(k)=
   \frac{e^{-i2R k}}  {\cos(2RK) - i\frac{k^2+K^2}{2 k K} \sin(2RK)} 
   \begin{pmatrix}
       1 & i\frac{2\mu}{\hbar^2} \frac{ V_0}{2 k K} \sin(2RK) \\
       i\frac{2\mu}{\hbar^2} \frac{ V_0}{2 k K} \sin(2RK)  & 1
   \end{pmatrix} \, .
\end{equation}
\end{widetext}

Up to now, we have considered the momentum variable $k$ to be a real number, or, equivalently, we have considered real energies. However, it is very useful to consider the transmission and reflection amplitudes as functions of the complex wavelength or complex energy. Doing so, we get the analytic extensions of $t(k)$ and $r(k)$, and then, the analytic extension of the $S$-matrix. At the common complex zeros of the denominator of $t$ and $r$, each element of the matrix $S$ will have a pole; we refer to these common poles of the matrix elements as poles of the $S$-matrix.

\begin{figure}[h!t]
\centering
	\includegraphics[width=0.7\linewidth,keepaspectratio]{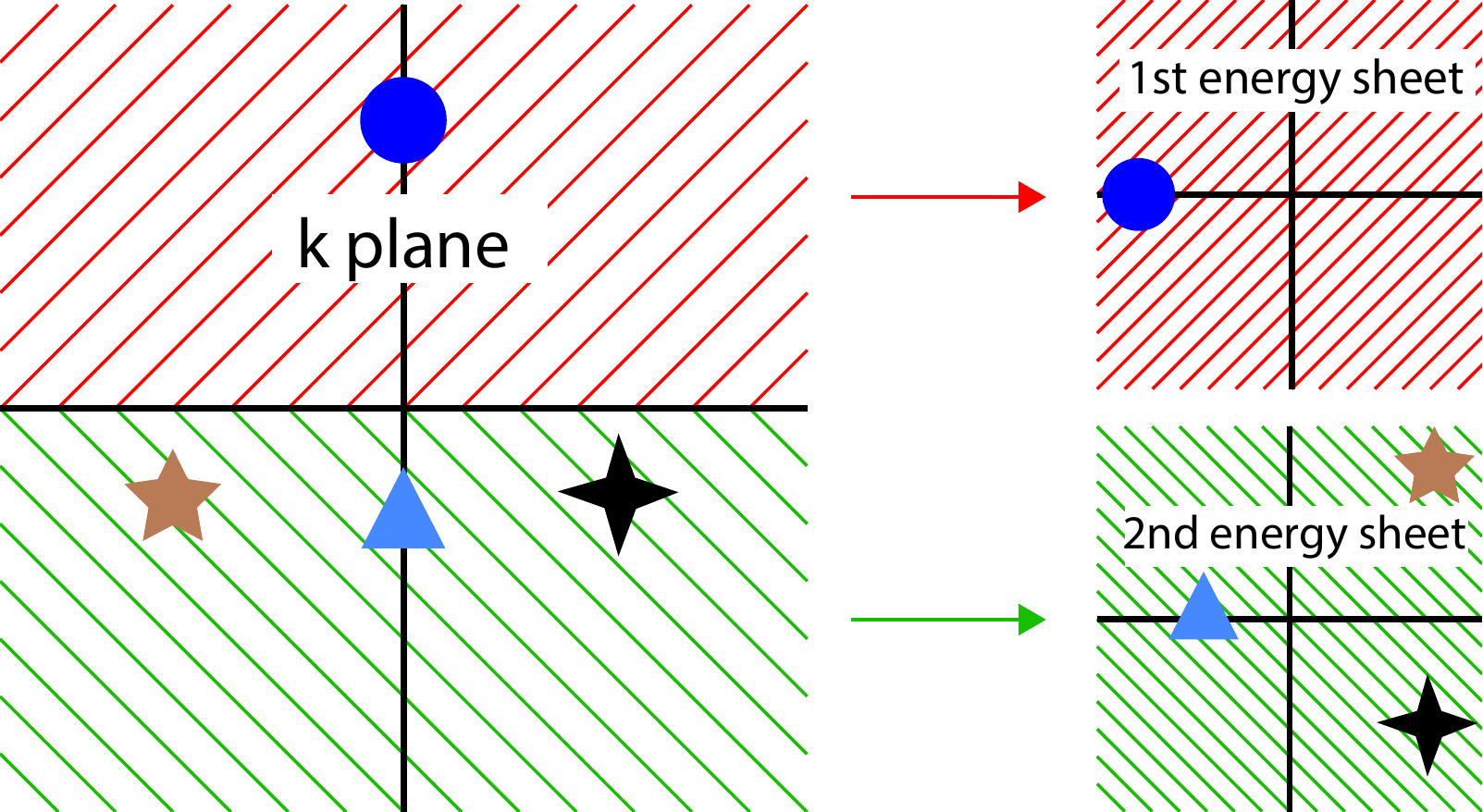}
	\caption{Mapping between the $k$ plane and the two energy sheets based on the dispersion relation $\varepsilon=\hbar^2k^2/2\mu $, Eq. (\ref{eq.kbound}).}
	\label{fig.riemann}
\end{figure}

The $S$-matrix in Eq. (\ref{eq.s}) connects the amplitudes of the wave function before and after the interaction, and thus it contains information about the system. In particular, its poles give the energies of the bound and unbound states of the system. For finite-range potentials (for further details the reader may consult Refs. \onlinecite{newtonScattering,1989Kukulin}), the poles of the $S$-matrix can be classified into three categories as shown in Fig. \ref{fig.riemann}: (i) bound states poles, which are sitting on the positive imaginary axis on the  complex wave number $k$-plane (filled circle); (ii) anti-bound states, which lie on the negative imaginary axis (filled triangle); and (iii) resonant poles, symmetrically placed with respect to the imaginary axis in the lower half plane (filled stars). Figure \ref{fig.riemann} also shows the corresponding positions of these poles in the complex energy plane. Notice that both bound and anti-bound states have real negative energies, since they are the square of a purely imaginary number. In Section \ref{sec.moving} we will show examples of all these poles.

To illustrate property (i) for the potential well let us search for the zeros of the common denominator of the elements of the $S$-matrix of Eq. (\ref{eq.smatrix}) in the complex $k$-plane
\begin{equation}\label{eq.deno}
	\cos(2RK) - i\frac{k^2+K^2}{2 k K} \sin(2RK)=0 \,.
\end{equation}
For the special case of bound states it is convenient to replace $k \rightarrow i \rho$, with $\rho \in \mathcal{R}^+$, then
\begin{equation}\label{eq.zeros}
	\cos(2RK) - \frac{K^2-\rho^2}{2 \rho K} \sin(2RK)=0
\end{equation}
with $K=\sqrt{\frac{2\mu}{\hbar^2}V_0 - \rho^2}$.  Notice that, if we use the dispersion relation $\varepsilon=\frac{\hbar^2}{2\mu} k^2$, with $k=i \rho$, the magnitude $K$ is the same as the one in Eq. (\ref{eq.kkappa}) for bound states. By using the identity $2\cot(2u)=\cot(u)-\tan(u)$ we obtain,
\begin{equation}\label{eq.cottan}
   \cot(RK) - \tan(RK) = \frac{K}{\rho} - \frac{\rho}{K} \, .
\end{equation}

The above equation has the structure $s-s^{-1}=f-f^{-1}$, with $s=\cot(RK)$ and $f=K/\rho$; which can be cast into the form $(s-f)(1+sf)=0$, with the solutions $s-f=0$ and $1+sf=0$. Using these solutions for Eq. (\ref{eq.cottan}) we get
\begin{align*}
   \rho &= K \tan(K R)\, , \\
   \rho &= -K \cot(RK) \, ,
\end{align*}
which are the equations we found in section \ref{sec.bound} for the bound solutions of the potential well. As a curiosity, notice that while in section \ref{sec.bound} we found these relations by searching for even and odd solutions independently, here we get both of them from a single condition, Eq. (\ref{eq.zeros}).

In Section \ref{sec.moving} we will illustrate numerically properties (ii) and (iii) by searching the zeros of Eq. (\ref{eq.zeros}) for anti-bound states and resonances. Now let us come back to the interpretation of the poles of the $S$-matrix in the complex energy plane. Using the dispersion relation we can map the upper and lower halves of the wave number $k$-plane onto two complex energy planes, see Fig. \ref{fig.riemann}. These two energy planes may be combined to form a single surface with two sheets called Riemann surface (see, for example chapter 2 of Ref. \onlinecite{2003Zill}). The bound states sitting on the positive imaginary $k$ axis are mapped to the negative real axis of the first sheet, called physical sheet (filled dots). The anti-bound states, found on the negative imaginary $k$ axis, are also mapped to the negatives real axis, but to the filled triangle on the second energy sheet, the so called nonphysical Riemann sheet. Finally, the symmetrical pair of resonances are mapped from the $k$-plane to the first and fourth quadrants, respectively, of the second Riemann sheet (filled stars). They are located symmetrically with respect to the real axis. 

Let us summarize the physical interpretation of the poles of the $S$-matrix. Bound states represent bound quantum systems, such as a nucleon (proton or neutron) in a nucleus (we will see an example in the first part of Sec. \ref{sec.deuteron}).  Anti-bound states and resonances represent unbound systems.\cite{1974Ohanian} Resonances show up as sharp maximums in the cross section as a function of the energy or as a rapid variation of the phase shift (we will see an example in the second part of Sec. \ref{sec.deuteron}). The physical interpretation of the anti-bound states is more elusive. We have mentioned that they have negative energies, like bound states, but they are 
not normalizable.  We found that they lie in the second Riemann sheet, which makes clear why this sheet is called nonphysical. An interpretation of an anti-bound or virtual state is that if the interaction were a bit stronger, the anti-bound state would become a bound state.\cite{1989Bohm,2019Moroz} In Sec. \ref{sec.moving} we will illustrate this transition for the potential well.

\subsection{Phase shift} \label{sec.phase}
The last picture we will use to illustrate resonances is the phase shift in a three-dimensional spherical potential well. As for the one-dimensional model, we write the Schr\"odinger equation in terms of the reduced mass and, for simplicity, we neglect the spin-orbit interaction:
\begin{equation} \label{eq.h3d}
   \left[ -\frac{\hbar^2}{2\mu} \nabla^2 
      + V(r) \right] \Psi(\boldsymbol{r})= \varepsilon \Psi(\boldsymbol{r}) \, .
\end{equation}
As usual, we write the three-dimensional Laplacian $\nabla^2$ as the sum of two terms, where $\boldsymbol{L}$ is the angular momentum operator:
$$\nabla^2=\frac{1}{r^2} \frac{\partial}{\partial r} \left( r^2 \frac{\partial}{\partial r} \right)
- \frac{\boldsymbol{L}^2/\hbar^2}{r^2}.$$

The eigenfunction $\Psi(\boldsymbol{r})$ can be written in terms of the spherical harmonics $Y_{lm}$ which are eigenfunctions of $\boldsymbol{L}^2$, with eigenvalues $l(l+1)\hbar^2$. By writing $\Psi(\boldsymbol{r})= r^{-1} u_l(r)  Y_{lm}(\theta,\phi)$ in Eq. (\ref{eq.h3d}) we obtain the following radial Schr\"odinger equation for $u_l(r)$: 
\begin{equation}
   \left[ 
   - \frac{\hbar^2}{2\mu}  \frac{d^2}{d r^2} 
   + \frac{l(l+1)\hbar^2}{2 \mu r^2}
    + V(r)
   \right] u_l(r)=\varepsilon u_l(r) \, .
\label{eq:schrodingerradial}
\end{equation}
where
$$
	V(r) =
	\begin{cases}
		-V_0 &  r \le  R\\
		0 & r > R
	\end{cases}$$
with $V_0>0$ and $R$ the strength and width, respectively, of the spherical potential well ($r \ge 0$).

The above equation looks very much like the one-dimensional Schr\"odinger equation with an extra term called centrifugal interaction. This term depends on the orbital angular momentum $l$, with $l=0,1,2,\cdots$. In order to introduce the phase shift we search for the scattering solutions $u_l(r)$ for $\varepsilon > 0$, that satisfy $u_l(r=0)=0$ (this condition makes $\Psi(\boldsymbol{r})$ be finite at zero), and $u_l$ and $u'_l$ continuous at $r=R$. Since the interaction does not couple solutions for different orbital angular momentum, we can write for each $l$
\begin{equation}
   u_l(r)=
   \begin{cases}
      A_l  j_l(K r) & 0\le r \le R \, \\
       h_l^-(k r)- S_l h_l^+(k r)  & r>R  
   \end{cases} \, ,
   \label{eq:arbitrarylsolution}
\end{equation}
with $k$ and $K$ as defined in Eq. (\ref{eq.kkappa}) for the one-dimensional potential well. The scattering wave function is written in terms of $j_l$ and $h_l^{\pm}$, i.e., the spherical Bessel and Hankel functions, respectively.\cite{abramowitz+stegun}  The coefficients $S_l$  for the partial waves $l$ correspond to the diagonal elements of the $S$-matrix. 
From Eq. (\ref{eq:arbitrarylsolution}) we have $|S_l|^2=1$, then we can write 
\begin{equation}\label{eq.sl}
	S_l=e^{2i\delta_l} \, ,
\end{equation}
with $\delta_l$ called the partial wave phase shift, for a reason that will be shown shortly. Requiring continuity of the wave function and its derivative at $r = R$ allows us to find the phase shift:\cite{newtonQP,newtonScattering} 
\begin{equation}
   \delta_l = 
       \arctan
       \left[
       \frac{k j_l'(k R) j_l(K R) - K j_l(k R) j_l'(K R)}
              {k \eta_l'(k R)j_l (K R) - K \eta_l(k R) j_l'(K R)}
      \right] \,  ,
   \label{eq:phaseshiftl}
\end{equation}
with $\eta_l$ the Neumann functions. We will illustrate the behavior of a phase shift as a function of the wave number $k$ in section \ref{sec.deuteron}.

In order to visualize that $\delta_l$ is associated to a shift in phase, let us replace the asymptotic   expressions of the Hankel functions 
\begin{align*}
	h_l^\pm(kr \rightarrow \infty) &=  e^{\pm i(kr-l \frac{\pi}{2})} \, , 
\end{align*}
in Eq. (\ref{eq:arbitrarylsolution}). 
Then, together with Eq. (\ref{eq.sl}) we get
\begin{equation}
   u_l(r\rightarrow \infty) = -2 i e^{i \delta_l} \sin(k r - \frac{\pi}{2} l + \delta_l) \,  ,
\end{equation}
which justifies the name of phase shift for $\delta_l$ (for more details about the physical interpretation of the phase shift, see for example, Sec. VIII.C in Ref. \onlinecite{cohenlibro} ).

As a final observation before moving to the applications, let us connect the properties of the amplitude $G(\omega)$ of Section \ref{sec.ho} with the cross section for a quantum resonance. The partial wave cross section $\sigma_l$ can be expressed in terms of the phase shift\cite{cohenlibro}:
\begin{equation}\label{eq.crosssect}
 	\sigma_l =\frac{4\pi}{k^2}  (2l+1) \sin^2(\delta_l) \; .
\end{equation}
For $\delta_l=\pi/2$, the cross section will be a maximum, and that maximum will be sharp if the phase shift rapidly changes around this value. These are the properties shown in Fig. \ref{fig:polos} for a resonance in the classical damped harmonic oscillator. Then, it is possible to write the phase shift as the right side of Eq.  (\ref{eq.bw2}) with $\gamma \rightarrow \Gamma/2$, and $\omega_0 \rightarrow \varepsilon_r$, where $\varepsilon_r$ and $\Gamma$ are the energy and width, respectively of the resonance,
\begin{equation}
   \delta_l=\arctan \left( \frac{\Gamma/2}{\varepsilon - \varepsilon_r}  \right) \, .
\end{equation}

By replacing this expression of the phase shift in Eq. (\ref{eq.crosssect}) we get
\begin{equation}
   \sigma_l =\frac{4\pi}{k^2}  (2l+1) 
   			\frac{\Gamma^2/4}{(\varepsilon-\varepsilon_r)^2-\Gamma^2/4} \, .
\end{equation}
This is the so called Breit-Wigner formula. Notice that this parametrization resembles that of the right side of Eq. (\ref{eq.bw}). The condition $\Gamma/\varepsilon_r \ll 1$ is equivalent to \textbf{a} sharp structure of the cross section. This ratio is like the one between the damping coefficient $\gamma$ and the resonant frequency $\omega_0$ we considered in Sec. \ref{sec.ho}, where we found that the sharp structure occurs when $\gamma/\omega_0 \ll 1$.

In summary, at the resonant energy, the phase shift takes the value $\pi/2$ and it rapidly changes in a small interval of $k$ or energy, while the cross section shows a sharp maximum around the resonant energy as a function of the energy. In Sec. \ref{sec:1d:application} we will see some examples of phase shift in the atomic nucleus.

\section{Applications}\label{sec:1d:application}
In this section we are going to illustrate the concepts developed in the previous one using some realistic parametrization for the potentials well . In Sec. \ref{sec.moving} we study the movement of the poles of the $S$-matrix and the behavior of the transmission coefficient for the one-dimensional potential well\cite{1996Sprung,2010Maheswari,2011Ahmed}. While in Sec. \ref{sec.deuteron} we use the three-dimensional potential well to illustrate the characteristic of the loosely bound state of the deuteron and the resonant phase shift of the nucleus $^5$He.

\subsection{Moving poles in the one-dimensional potential well} \label{sec.moving}
The goal of this section is to illustrate the three categories of poles of the $S$-matrix presented in Section \ref{sec.smatrix}. We also show the transitions between bound, anti-bound and resonant states. We will calculate the poles of the $S$-matrix Eq. (\ref{eq.smatrix}) as a function of the potential strength $V_0$ with $R=2.1$ fm and $2 \mu/\hbar^2=0.024 \, \rm{MeV}^{-1}\, \rm{fm}^{-2}$.

\begin{figure}[h!t]
\centering
	\includegraphics[width=0.8\linewidth,keepaspectratio]{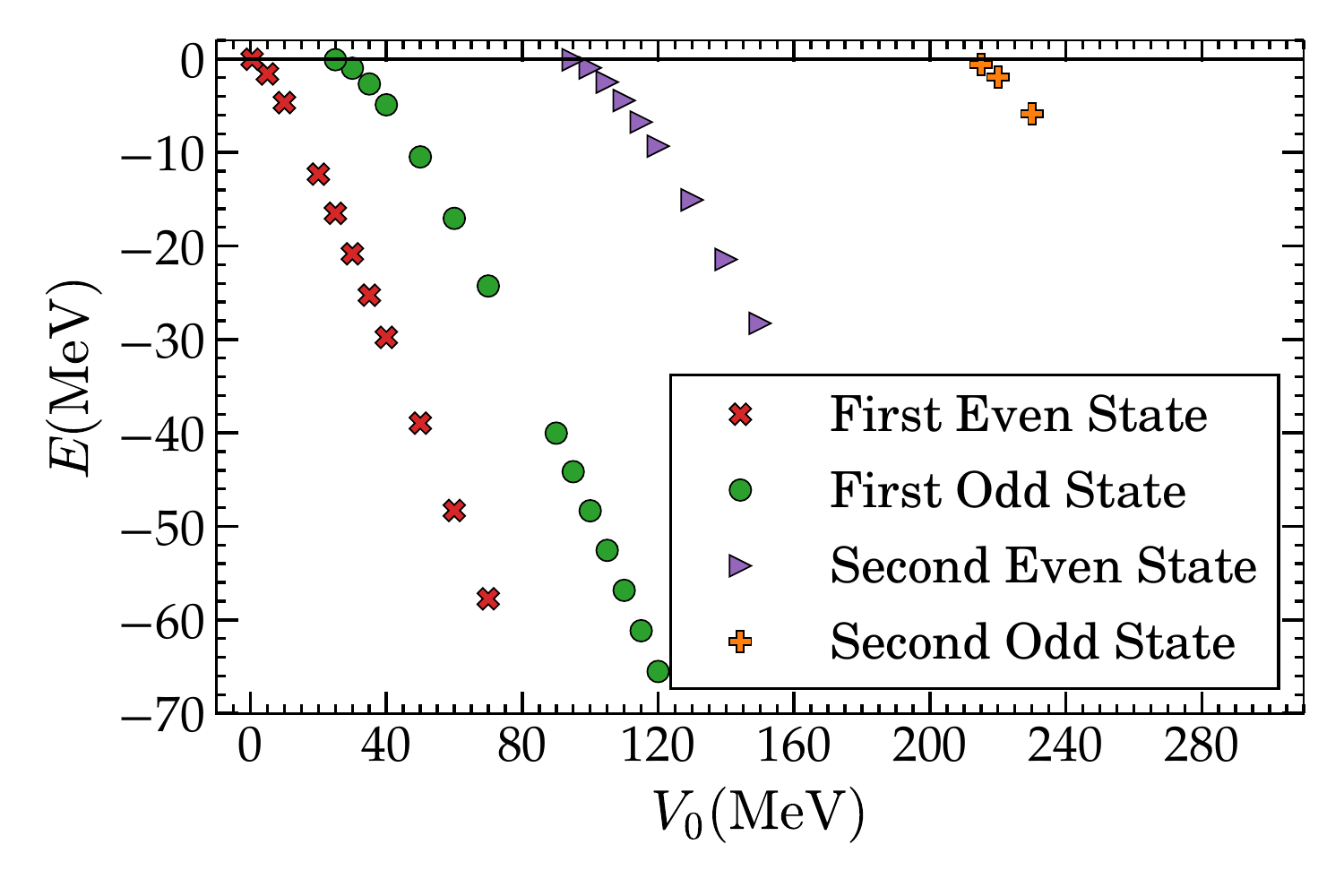}
	\caption{Evolution of the first two even and two odd bound states in the one-dimensional square well $R=2.1$ fm and $2 \mu / \hbar^2=0.024$ MeV$^{-1}$ fm$^{-2}$, for $0<V_0 < 250$ MeV.}
	\label{fig:firststatesmove}
\end{figure}

We start by calculating the first four bound states using Eqs. (\ref{eq.even}) and (\ref{eq.odd}), which are shown in Fig. \ref{fig:firststatesmove} as a function of $V_0$. We see that for any value of the strength there is always a bound state with even parity; as the strength increases this state becomes more bound. A new bound state, with odd parity,  appears for $V_0 \sim 23.3$ MeV. As the strength increases, more states appear with interleaved parity.

\begin{figure}[h!t]
	\centering
	\includegraphics[width=0.8\linewidth,keepaspectratio]{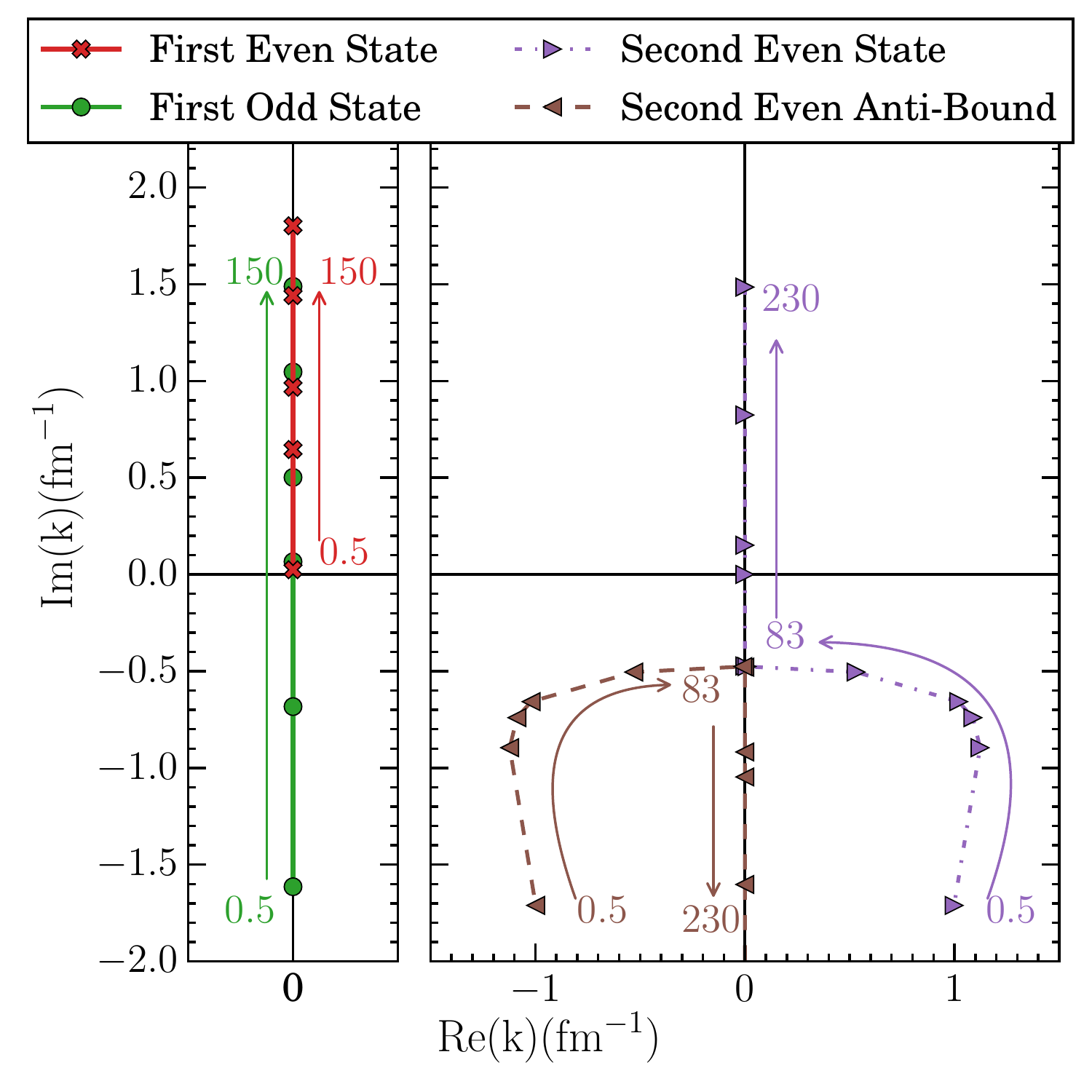}
	\caption{Moving of the first few poles of the $S$-matrix Eq. (\ref{eq.smatrix}) parameterized in $V_0$ for the one-dimensional potential well, with $R=2.1$ fm and $2 \mu / \hbar^2=0.024$ MeV$^{-1}$ fm$^{-2}$. The arrows indicate the direction of increasing $V_0$. The numbers next to the arrows are values of $V_0$ in MeV. (Left) Trajectories of the first even and odd states. (Right) Trajectories of the first pair of poles.}
	\label{fig:polesmove}
\end{figure}

Next, we study the locations of the first three states of Fig. \ref{fig:firststatesmove} in the complex $k$-plane by finding the zeros of Eq. (\ref{eq.deno}) for different values of $V_0$. The left panel of Fig. \ref{fig:polesmove} shows the movement of the first even state (x's) and first odd state (circles), while the right panel shows the evolution of the second even state (triangles).  The first even bound state starts with $k = 0$ and moves along the positive imaginary axis as $V_0$ increases, implying a state that becomes increasingly more bound as the well gets deeper. The first odd state, however, starts at a large negative value along the imaginary axis. It moves upwards along the negative imaginary axis as $V_0$ increases, reaching the origin and becoming a bound state at $V_0 =  23.29$ MeV, consistent with Fig. \ref{fig:firststatesmove}. 

The second even state is missing as a bound state in Fig. \ref{fig:firststatesmove} for small values of the strength. In order to find its evolution we search for complex zeros of Eq. (\ref{eq.deno}) with non zero real part for $V_0$ small, i.e., we search for resonances. The right panel of Fig. \ref{fig:polesmove} shows the evolution of a pair of symmetric poles ($n=3$ of Ref. \cite{1996Sprung}) which moves upwards in the third and fourth quadrant of the $k$-plane. 
This pair of poles merges on the negative imaginary axis at $k \simeq  -0.5 i$ fm$^{-1}$ ($V_0 \simeq 83$ MeV) into a second-order pole. 
Afterward, one of the poles moves up and the other moves down, both along the imaginary axis. Eventually, as the strength $V_0$ keeps growing, one pole moves to positive imaginary values of $k$  and becomes the second even bound state (right triangle) that appears in Fig. \ref{fig:firststatesmove} for $V_0 \gtrsim 83$ MeV.

In summary, we have shown in Fig. \ref{fig:polesmove} how the different kinds of poles of the $S$-matrix evolve as a function of the strength. Let us remark on a detail which will be relevant for the next section: for a one-dimensional potential well there always exists at least one bound state for any value of $V_0$; this is not the case for a three-dimensional well.

\begin{figure}[h!t]
\centering
   \includegraphics[width=0.8\linewidth,keepaspectratio]{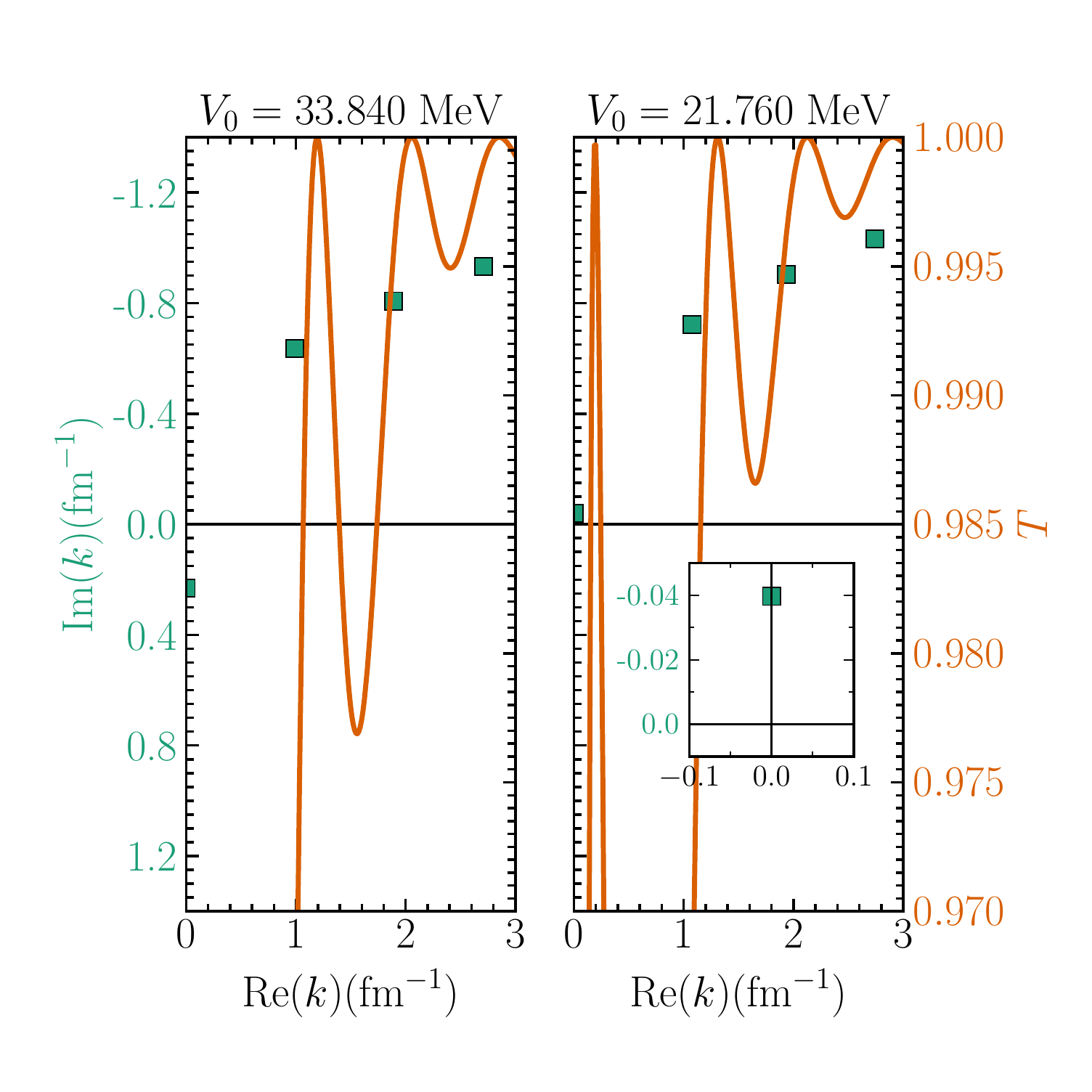}
   \caption{ Transmission coefficient (continuum curve) and complex poles (filled squares) of the transmission coefficient Eq.  (\ref{eq:transcoef}) for $V_0=33.84$ MeV (left) and $V_0=21.76$ MeV (right). The inset shows the position of the anti-bound. }
   \label{fig:tcoef}
\end{figure}

Let us now study the behavior of the transmission coefficient $T$, Eq. (\ref{eq:transcoef}), as a function of the real wave number $k$, for two values of the strength, $V_0=33.84$ and $21.76$ MeV (these figures corresponds to the bound and anti-bound strengths for the deuteron of Sec. \ref{sec.deuteron}). 
The left and right panels of Fig. \ref{fig:tcoef} show $T$ for $V_0=33.84$ and $21.76$ MeV, respectively, with the scale on the right. Both curves show the characteristic behavior of $T$, with some resonant peaks for which $T=1$; these peaks widen as $k$ increases. Notice that the right figure has a very narrow peak close to $k=0$. 

In Fig. \ref{fig:tcoef} we also show the poles of the $S$-matrix as functions of the complex $k$ as filled squares, with the scale on the left side. For a better visualization of the position of the poles with respect to the peaks of $T$, the scale of the left axis is inverted. We observe that the real part of the poles are close to the position of the peaks, and that the agreement deteriorates as the imaginary part of the pole increases. We also observe that as the peaks widens, the imaginary part of the poles increase. 
The broadening of the peaks with the increment of the imaginary part of the poles is reminiscent of the one we found for $|G|^2$ in Fig. \ref{fig:evolucion} as the damping coefficient increases. 
The pole on the positive imaginary axis in the left figure corresponds to a bound state, while the pole on the negative imaginary axis on the right figure corresponds to an anti-bound state.

\subsection{Loosely-bound and resonant states in real nuclei} \label{sec.deuteron}
In this section we will connect the bound and anti-bound states found in Fig. \ref{fig:tcoef} with the corresponding states of the deuteron. We also will show the behavior of the phase shift for the nucleus $^5$He.

First, let us consider the one-dimensional potential well as an approximation for the interaction in the deuteron, the nucleus of the heavy hydrogen atom, $^2$H. It can be solved by reducing the two-body proton-neutron system to a one-body problem with reduced mass $\mu$, with $2 \mu/\hbar^2=0.024\,  \rm{MeV}^{-1} \, \rm{fm}^{-2}$. The strength $V_0$ measures the interaction between the proton and the neutron. For the parameter $R$ of the well we take the experimental radius of the deuteron $R=2.1$ fm (p. 40 in Ref. \onlinecite{2007Bertulani}). 
Notice that these are the parameters used in the previous section to study the positions of the poles of the $S$-matrix. The experimental binding energy of the deuteron is $\varepsilon=-2.225$ MeV \cite{nndc}.

In our one-dimensional potential well the first bound state is an even state then, in order to describe the deuteron we have to consider the second bound state of the one-dimensional potential well, i.e. the first odd state. Taken for the strength the value $V_0=33.84$ MeV we reproduce the ground state energy $-2.225$ MeV of the deuteron.
It appear as a bound state pole in the left panel of Fig. \ref{fig:tcoef}, at the value $k=+i\, 0.231$ fm$^{-1}$. 
The deuteron has an unbound state at $\varepsilon=-0.066$ MeV, \cite{1989Kukulin} 
which is found, in our one-dimensional potential well for $V_0=21.76$ MeV. This state appears as an anti-bound pole on the right panel in Fig. \ref{fig:tcoef}, at the value $k=-i\, 0.040 $ fm$^{-1}$. 
Different values of $V_0$ were needed to describe the bound and anti-bound states of the deuteron because of our oversimplified model. A more realistic description of the deuteron would require a more complex interaction. 

\begin{figure}[h!t]
	\centering
	\includegraphics[width=0.8\linewidth,keepaspectratio]{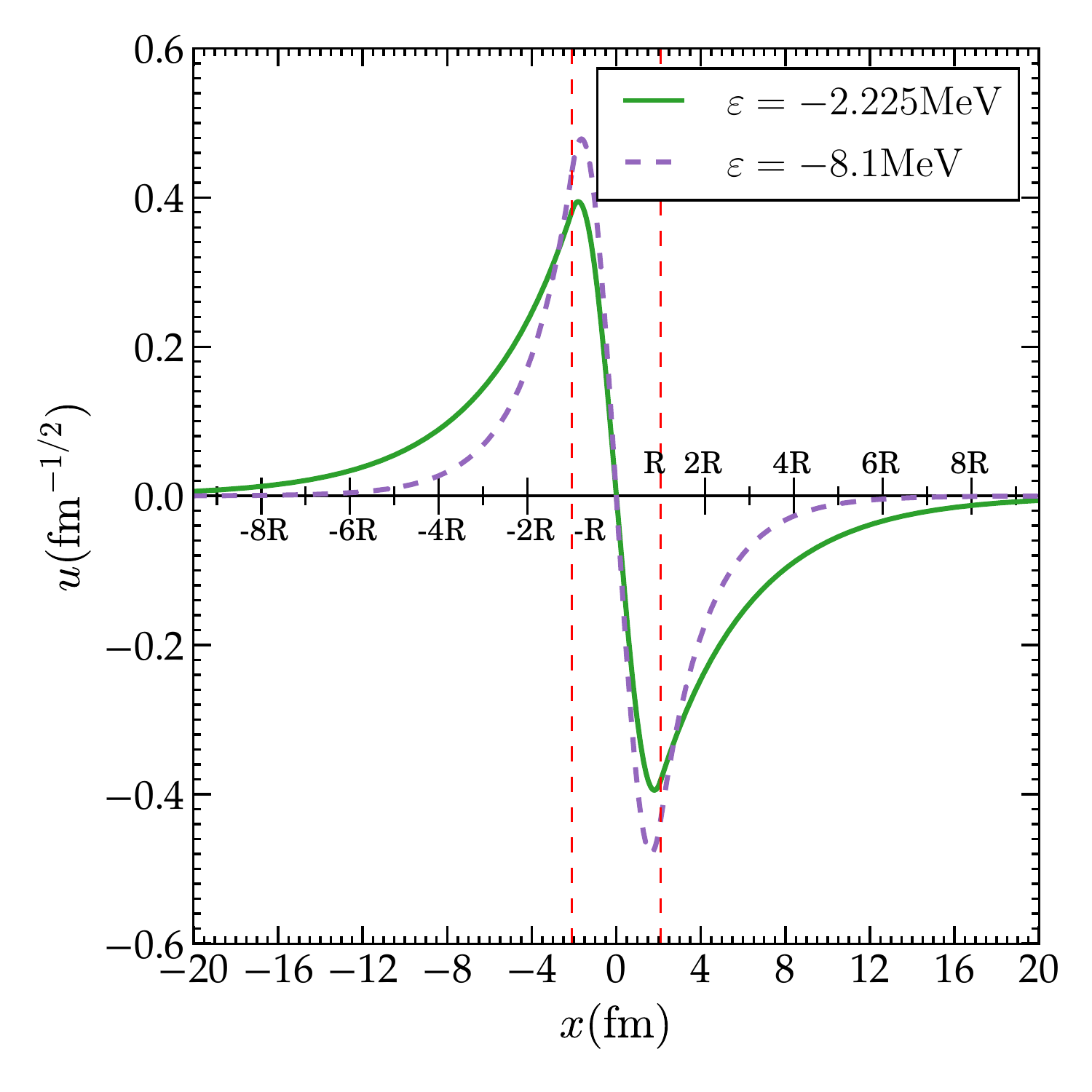}
	\caption{Comparison of the wave function of the deuteron (solid line) with that of a typical bound state (dashed line). The vertical dashed lines delimit the radius $R=2.1$ fm.}
	\label{fig:wavefunct}
\end{figure}

The deuteron is an example of the so called loosely bound nuclei. The characteristic of these nuclei is that the wave function extends well beyond the radius $R$. In order to have an idea about this property, we artificially increases the strength to $V_0=46$ MeV; the energy of this state is $\varepsilon=-8.1$ MeV. Figure \ref{fig:wavefunct} shows the wave function of this state with that of the deuteron. We observe that the wave function of the state with energy $\varepsilon=-8.1$ MeV decay much faster than that of the deuteron, with energy $\varepsilon=-2.225$ MeV.

\begin{figure}[h!t]
	\centering
	\includegraphics[width=0.95\linewidth,keepaspectratio]{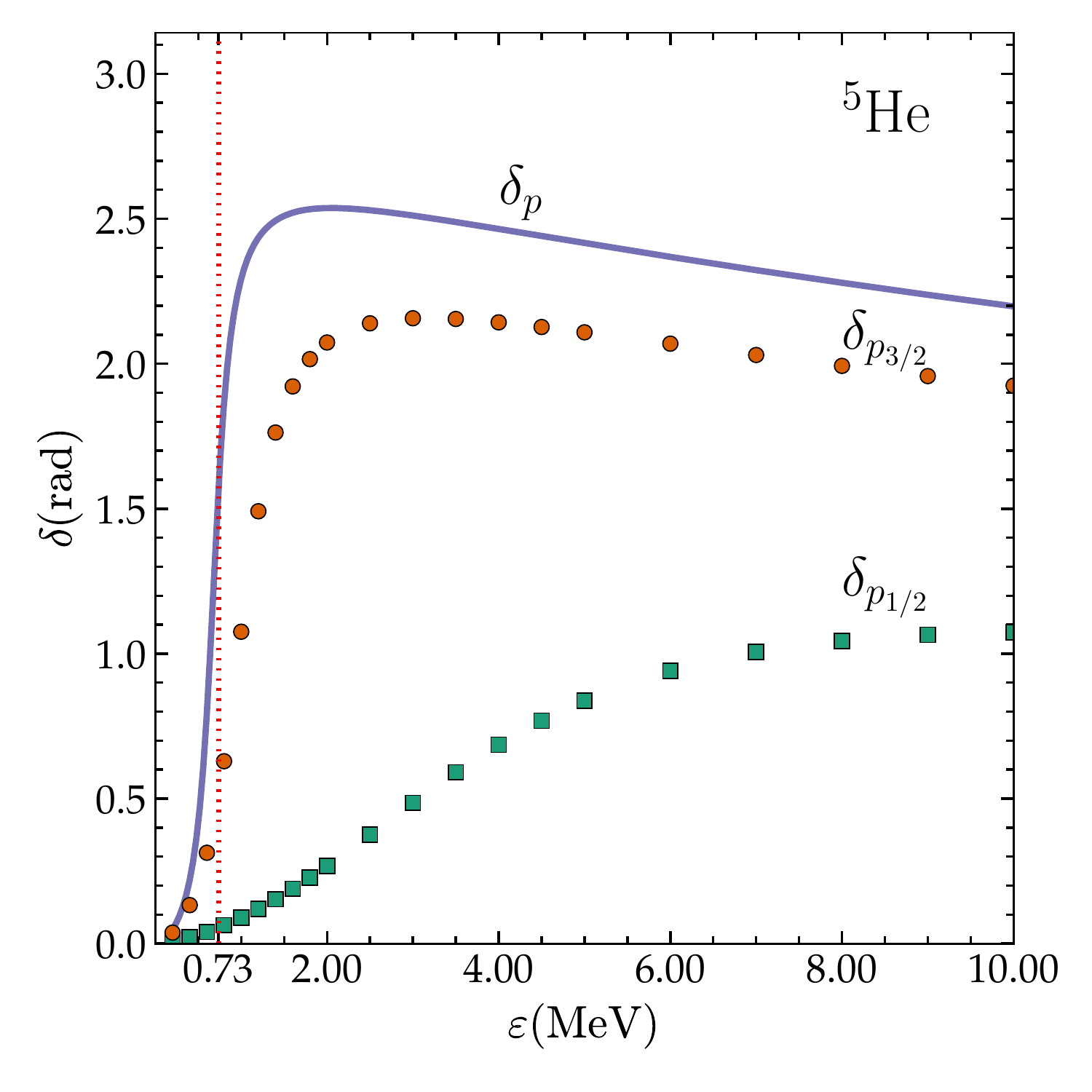}
	\caption{Calculated phase shift $\delta_{p}$ for $^5$He in the three-dimensional potential well and the corresponding experimental phase shifts $\delta_{p_{3/2}}$ and $\delta_{p_{1/2}}$ \cite{1977Bond}. The vertical dotted line shows the experimental resonant energy $\varepsilon_{3/2^-}=0.735$MeV.}
	\label{fig:phaseshift}
\end{figure}

As a final example of real nucleus, let us consider the $^5$He isotope. Experimentally, it is found that the separation energy of this nucleus is negative\cite{nndc}, meaning that it is unbound. From the nuclear shell model the $^5$He nucleus is modeled as a two-body system, a $^4$He core plus a neutron with intrinsic spin $s=1/2$. The spin $\boldsymbol{s}$ couples to the orbital angular momentum $\boldsymbol{l}$ to give the total angular momentum $\boldsymbol{j}= \boldsymbol{l} +\boldsymbol{s}$. In this picture the valence neutron may be in the $p_{3/2}$ or $p_{1/2}$ (with  $p$ indicating the orbital angular momentum $l=1$ and $j=3/2$ or $j=1/2$). The reduced mass of the $^4$He plus neutron system yields $2\mu/\hbar^2=0.0385$ MeV$^{-1}$fm$^{-2}$, while the value for the parameter $R$ was set equal to $2.15$ fm. 
Since we are not considering the spin-orbit interaction, the states $p_{3/2}$ and $p_{1/2}$ are degenerate in a single $p$ state. The strength of the potential well $V_0=53.18$ MeV was chosen to have $\delta_{p}=\pi / 2$ at the experimental \cite{nndc} resonant energy $\varepsilon_{3/2^-}=0.735$ MeV. Using Eq. (\ref{eq:phaseshiftl}) we calculate the phase shift $\delta_p$ shown by a solid line in Fig. \ref{fig:phaseshift}. The profile of the calculated phase shift $\delta_p$ is reminiscent of the function $Arg(G)$ of the damped harmonic oscillator of Fig. \ref{fig:polos}. From the experimentally-measured cross-section as a function of the energy one may extract the phase shift for the different quantum numbers. In Fig. \ref{fig:phaseshift} we show the experimental phase shift for $p_{3/2}$ and $p_{1/2}$ (data from Ref. \onlinecite{1977Bond}). We observe that the calculated phase shift $\delta_p$ in our simplified model follows closer the shape of the experimental phase shift $\delta_{p_{3/2}}$, i.e. the resonant behaviour of the $^5$He nucleus. Better agreement with the experimental data would require the inclusion of the spin-orbit interaction.

The condition $\Gamma/\varepsilon_r \ll 1$ is usually used to characterize an observable resonance \cite{1989Bohm,1989Kukulin}, with $\varepsilon_r$ the energy of the resonance and $\Gamma$ its width. From Fig. \ref{fig:phaseshift} we can calculate the width \cite{2001Bohm} 
$$\Gamma=2 \left( \frac{d \delta}{d \varepsilon} \right)^{-1}_{\varepsilon_r}=0.373~\rm{ MeV} \, .$$ 
Then, by taking the ratio between the width $\Gamma$ and the experimental resonant energy $\varepsilon_r=0.735$ MeV \cite{nndc} for the $^5$He, we get $\Gamma / \varepsilon_r =0.51$ MeV. Since this ratio is not as small as for the harmonic oscillator, the jump in Fig. \ref{fig:phaseshift} is somewhat less pronounced that in Fig. \ref{fig:polos}.

\section{Summary} \label{sec.conclusions}
We started by reviewing the behavior of the amplitude and phase of the classical damped harmonic oscillator near resonance. Then we applied the usual matching procedure to find the bound and scattering solutions of the one-dimensional well potential. After that we defined the $S$-matrix for the one dimensional potential well and introduced the its analytic extension to the complex wave number plane. We classified the poles of the scattering matrix and made the connection between its poles in the complex wave number and energy planes. Finally we introduced the phase shift in the three-dimensional spherical potential well. 

In the applications we showed the movement of the first few poles of scattering matrix for the one-dimensional potential well in the complex wave number plane as a function of the strength of the potential. We showed how the first state is always present for any value of the strength and how the second state becomes a bound state starting as an anti-bound state. The movement of these two states in the energy plane is as follow: the first state moves always on the physical (first) Riemann sheet along the negative real axis from zero towards negative infinity as the strength increases, while the second state also moves along the negative real axis but it starts on the nonphysical (second) Riemann sheet and then continues along the negative real axis on the physical Riemann sheet. The third bound state has an anti-bound partner; both these states started as resonances from small values of the strength and they merged on the negative imaginary axis, as it was found for the classical harmonic oscillator. Also, for the one-dimensional potential well, we illustrated the relation between the isolated poles of the scattering matrix and the maximums of the transmission coefficient. We found that the agreement departed as the imaginary part of the poles widens. We showed that the anti-bound state showed up as a narrow peak in the transmission coefficient very close to the origin. 
 
By using the one-dimensional potential well we compared the loosely bound state wave function of the deuteron with a more tightly bound state and showed how the former extends much more in the space than the latter. As an example of an experimentally observable resonance in nuclear physics, we used the three-dimensional potential well as an approximate model to illustrated the unbound $^5$He nucleus. We found the usual rapid variation of the phase shift that characterizes a resonance. 

The use of the potential well in one and three dimensions allowed us to build toy models of real nuclei which were analytically solved. The concepts and equations presented in this article will allow advanced students and instructors to understand resonant properties in real nuclei by themselves, varying the parameters that define the Schr\"{o}dinger equation.

\begin{acknowledgments}
This work has been supported by the University of Rosario ING588, Argentina.
\end{acknowledgments}

%

\end{document}